%Paper: hep-th/9305087
%From: Stefan=Vandoren%TF%FYS@cc3.kuleuven.ac.be
%Date: Wed, 19 May 93 10:29:28 CET

\magnification = 1200
\hsize = 15 truecm
\vsize = 23 truecm
\baselineskip 20 truept
\voffset = -0.5 truecm
\parindent = 1 cm
\overfullrule = 0pt
\def\u{1\kern -0.7em \hbox {1}}
\count0 = 0
\footline={\hfil}
\null
{\hskip 8 cm  {Preprint-KUL-TF-93/15}}
\vskip 1 truecm

\centerline
{\bf The Batalin-Vilkovisky formalism on fermionic K\"ahler manifolds}

\vskip 3 truecm

\centerline{\bf  S. Aoyama $^1$ and S. Vandoren $^2$}
\smallskip

\vskip 0.5 cm
\centerline{\bf Instituut voor Theoretische Fysica}
\smallskip
\centerline{\bf Katholieke Universiteit Leuven}
\smallskip
\centerline{\bf Celestijnenlaan 200D}
\smallskip
\centerline{\bf B-3001 Leuven, Belgium}

\vskip 2.5 truecm

\noindent
{\bf Abstract}

We show that the K\"ahler structure can be naturally incorporated in the
Batalin-Vilkovisky formalism. The phase space of the BV formalism becomes a
fermionic K\"ahler manifold. By introducing an isometry we explicitly
construct the  fermionic irreducible hermitian symmetric space. We then give
some solutions of the master equation in the BV formalism.

\vskip 3 cm
\noindent $^1$ E-mail : Shogo\%tf\%fys@cc3.kuleuven.ac.be

\noindent $^2$ E-mail : Stefan\%tf\%fys@cc3.kuleuven.ac.be
\vfill\eject

\footline={\hss\tenrm\folio\hss}

\noindent
{\bf 1.~~Introduction}

The anti-bracket formalism  already appeared in the beginning of the history
of the BRST quantization. It was a useful tool to discuss renormalization of
the non-abelian gauge theory in the BRST quantization$^{[1]}$. The formalism
has been fairly elaborated by Batalin and Vilkovisky. It became a viable
formalism for the BRST
quantization of general gauge theories$^{[2]}$. For instance, it has been
successfully applied$^{[3]}$  to gauge theories with open algebrae$^{[4]}$.
The geometrical meaning of this
 BV formalism has been considerably clarified by Witten$^{[5]}$. More recently
 the BV formalism has been set up on a curved supermanifold of fields and
 anti-fields with a fermionic symplectic structure$^{[6]}$. It has been
 applied to study
 quantization of the string field theory$^{[7,8]}$. The application went far
 beyond the
original motivation of the BRST quantization. As such an example
we would also like to mention the work by Verlinde$^{[9]}$.

 In whatever circumstance it is used,  the ultimate goal of the BV formalism
 is to determine the fermionic symplectic structure of the supermanifold and
 solve the master equation. Therefore it is important to understand the
 geometry of the fermionic
symplectic structure.

 In this note we introduce the K\"ahler structure to the supermanifold, and
 show that the symplectic structure is reduced to that given by a fermionic
 K\"ahler 2-form $\omega$. The K\"ahler potential, which is the hallmark of
 such a supermanifold, is
then fermionic.  Secondly we introduce an isometry to the supermanifold. It
is done similarly to the case of the bosonic K\"ahler manifold$^{[10]}$. The
only complication is due to  sign factors coming from ordering  fermionic
coordinates.
The isometry is realized by the Killing vectors.
They are given by
a set of real potentials, which we call the Killing potentials. Our main
message in this regard is that for a class of supermanifolds the fermionic
K\"ahler 2-form $\omega$  can be explicitly constructed out of the Killing
vectors
of the (bosonic) irreducible hermitian symmetric space$^{[11]}$.
The supermanifold with a symplectic structure given by this 2-form is
called the fermionic irreducible hermitian symmetric space.
Finally we are interested in solving the master equation with the fermionic
symplectic structure given above. In the first place it is solved for the
fermionic CP$^1$ space by assuming that  the coordinates of the space are
space-time independent.
If they are not, the master equation requires some appropriate quantum
consideration$^{[12]}$. We are not going to be involved in this problem. We
only investigate the classical master equation, which is still  important to
study  before quantization.
Since it is a non-linear functional equation, many solutions are expected in
principle. We find one solution for the fermionic irreducible hermitian
symmetric space . It is given in terms of the Killing potentials.

 Fermionic K\"ahler manifolds in the BV formalism have been discussed in
 the recent paper$^{[13]}$. However they did not study the isometry to
 compute the metric and the Killing potentials of the manifold. These
 quantities are important to find solutions
of the master equation, as we will see.

\vskip 1 cm

\noindent
{\bf 2.~~The BV formalism}

  Let us start with a short review on the BV formalism. Consider a 2$D$
manifold parametrized by real coordinates $y^i = (x^1,x^2,\cdots,x^D,
\xi^1,\xi^2,\cdots,\xi^D) $ with $x$'s and $\xi$'s bosonic and fermionic
respectively. Suppose
that it has a symplectic structure given by a non-degenerate 2-form
$$
\omega = dy^j \wedge dy^i \omega_{ij}   ,               \eqno (2.1)
$$
which is closed
$$
d\omega = 0.          \eqno (2.2)
$$
These equations read in components
$$
(-)^{ik}  \partial_i \omega_{jk} + (-)^{ji} \partial_j \omega_{ki} +
(-)^{kj}\partial_k \omega_{ij} = 0 ,    \eqno (2.3)
$$
$$
\omega_{ij}  = -(-)^{ij}\omega_{ji}.  \eqno (2.4)
$$
Here  one should understand the short-hand notation for the grassmannian parity
of the coordinates
$\varepsilon (y^i) = i $ in the sign factor. By this notation we have
$\varepsilon (\omega_{ij}) = i + j + 1$. We define the anti-bracket by $$
\{A,B\} = A\overleftarrow \partial_i \omega^{ij} \partial_j B,
\eqno (2.5)$$
in which $\omega^{ij}$ is the inverse matrix of $\omega_{ij}$ such that
$$
\omega_{ij} \omega^{jk} = \omega^{kj}\omega_{ji} = \delta^k_i. \eqno (2.6)
$$
Note that the right-derivative $\overleftarrow \partial_i$ is related with
the right-one by
$$
A\overleftarrow \partial_i = (-)^{i(\varepsilon (A) + 1)} \partial_i A  .
$$
In terms of $\omega^{ij}$ eqs (2.3) and (2.4) become respectively
$$
(-)^{(i+1)(k+1)} \omega^{il}  \partial_l \omega^{jk} +
(-)^{(j+1)(i+1)} \omega^{jl}\partial_l \omega^{ki} +
(-)^{(k+1)(j+1)} \omega^{kl}\partial_l \omega^{ij}  = 0, \eqno (2.7)
$$
$$
\omega^{ij}  = -(-)^{(i+1)(j+1)} \omega^{ji}  .   \eqno (2.8)
$$
Owing to eq. (2.7) the anti-bracket (2.5) satisfies the Jacobi identity.

\vskip 1 cm

We define a second order differential operator by
$$
\Delta \equiv {1 \over \rho }(-)^i \partial_i(\rho \omega^{ij}\partial_j),
\eqno (2.9)
$$
with a bosonic function $\rho$. It satisfies the following properties:
$$
\eqalign{
\Delta\{A,B\} & = \{\Delta A,B\} + (-)^{\varepsilon (A)+1}\{A,\Delta B\},
\cr
\Delta (AB) & = \Delta A \cdot B + (-)^{\varepsilon (A)} A\Delta B  +
(-)^{\varepsilon (A) } \{A,B \}. \cr }
$$
Moreover the operator $\Delta$ is nilpotent $\Delta^2 = 0$ if $\rho$ obeys
the equation
$$
\Delta [ {1 \over \rho}(-)^i\partial_i (\rho \omega^{ij}) ] = 0  .
\eqno (2.11)
$$
Finally the master equation in the BV formalism is given by
$$
\Delta e^S = 0  , \quad \quad {\rm or}    \quad \quad
\Delta S + {1\over 2}\{S,S\} = 0 .    \eqno (2.12)
$$

\vskip 2 cm

\noindent
{\bf 3.~~The fermionic K\"ahler geometry}

 So far we have discussed the symplectic form of the $2D$ supermanifold.
When  $D=2d$, the metric of the manifold $\gamma_{ij} $ may be defined by
$$
\gamma_{ij} = \omega_{ik}J^k_{\ j}                   \eqno (3.1)
$$
in which
$$
J^k_{\ j} =\left(
\matrix{ 0  & \u & 0 & 0  \cr
         -\u & 0 & 0 & 0  \cr
         0  & 0 & 0 & \u  \cr
         0  & 0 & -\u & 0 \cr}
\right)
$$
with the $d \times d$ unit matrix $\u$.
We shall now impose the condition on $\omega_{ij}$
$$
\omega_{kl}J^k_{\ i}J^l_{\ j}  = \omega_{ij}       \eqno (3.2)
$$
or equivalently
$$
\gamma_{kl}J^k_{\ i}J^l_{\ j}  = \gamma_{ij}       \eqno (3.3)
$$
Since  $J^k_{\ j}$ is a bosonic matrix, the metric $\gamma_{ij}$ is fermionic,
$ \varepsilon (\gamma_{ij}) = i + j + 1$. From eqs (2.4), (2.6) and (2.8) we
have
$$
\eqalign{
\gamma_{ij} & = (-)^{ij} \gamma_{ji},  \cr
\gamma_{ij}\gamma^{jk} & = \gamma^{kj}\gamma_{ji} = \delta^k_i,
\cr
\gamma^{ij} & = (-)^{(i+1)(j+1)}\gamma^{ji}. \cr} \eqno (3.4)
$$
We also find the relation
$$
\gamma_{ij}\omega^{jk}\gamma_{kl} = -\omega_{il}.  \eqno (3.5)
$$
The affine connection may be defined by postulating
$$
D_k \gamma_{ij} \equiv \partial_k \gamma_{ij} - \Gamma^l_{ki}\gamma_{lj}
- (-)^{ij}\Gamma^l_{kj}\gamma_{li} = 0.
$$
By solving this we obtain the connection
$$
\Gamma^k_{ij} = {1 \over 2}[\partial_i \gamma_{jl} + (-)^{ij}\partial_j
\gamma_{il} - \gamma_{ij}\overleftarrow \partial_l ]\gamma^{lk},    \eqno (3.6)
$$
 which is bosonic, $\varepsilon (\Gamma^k_{ij}) = i+j+k$.

\vskip 1 cm

We shall go to the complex coordinate basis
$
y^i \rightarrow ({\bf z}^a, \overline {\bf z}^{\underline a})
$
with
$$
{\bf z}^a  = (z^\alpha,  \zeta^\alpha),  \quad \quad
\overline {\bf z}^{\underline a}  = (\overline z^{\underline \alpha},
\overline \zeta^{\underline \alpha}),  \quad \quad
\alpha  = 1,2,\cdots,d ,
$$
defined by
$$
z^{\alpha} = x^\alpha + i x^{d+\alpha}, \quad  \quad
\zeta^\alpha = \xi^\alpha + i \xi^{d+\alpha},  \quad \quad {\rm c.c.}\quad .
$$ The condition (3.2)  or (3.3) reduces to
$$
\eqalign{
\omega_{ab} & = \omega_{\underline a \underline b} = 0, \quad \quad \quad
\gamma_{ab}  = \gamma_{\underline a \underline b} = 0,  \cr
\omega_{a \underline b} & = i \gamma_{a \underline b}, \quad \quad \quad
\omega_{\underline a b} = -i \gamma_{\underline a b},  \cr}  \eqno (3.7)
$$
together with \footnote \dag {Complex conjugation of fermion products is
chosen as $(\zeta \eta)^\ast = \overline \zeta \overline \eta$.} $$
\eqalign{
\omega_{a \underline b} & = -(-)^{ab}\omega_{\underline b a},
\quad \quad \quad  \omega_{a \underline b}^\ast = \omega_{\underline a b},
\cr
\gamma_{a \underline b} & = (-)^{ab}\gamma_{\underline b a},
\quad \quad \quad  \gamma_{a \underline b}^\ast = \gamma_{\underline a b}.
\cr}  \eqno (3.8)
$$
By means of these equations the symplectic form (2.1) takes the K\"ahler
2-form
$$
\omega = 2i d \overline {\bf z}^{\underline b}\wedge d {\bf z}^a \
\gamma_{a \underline b}.   \eqno (3.9)
$$
Then eq. (2.2) or equivalently (2.3) is solved by
$$
\gamma_{a \underline b} = \partial_a \partial_{\underline b} K, \eqno (3.10)
$$
in which $K$ is a fermionic K\"ahler potential.
Thus the supermanifold acquires a fermionic K\"ahler geometry as the
consequence of eqs (2.4)  and (3.2).
The affine connection (3.6) is simplified as
$$
\Gamma^c_{ab}  = \partial_a \gamma_{b \underline d}\cdot \gamma^
{\underline d c},    \quad \quad
\Gamma^{\underline c}_{\underline a \underline b}  = \partial_{\underline a}
\gamma_{\underline b d}\cdot \gamma^{d \underline c} ,
$$
and all the other components are vanishing.
The covariant derivative of a holomorphic vector is defined by
$$
D_a A_b \equiv \partial_a A_b - \Gamma^c_{ab}A_c,  \quad \quad
D_a A^b \equiv \partial_a A^b + (-)^{a \varepsilon (A)}A^c\Gamma^b_{ca}.
$$

\vskip 1 cm

\noindent
{\bf 4.~~The isometry}

The fermionic K\"ahler manifold admits an isometry. It is realized by a set of
Killing vectors $V^{Ai}(y), A = 1,2,\cdots,N$, with  $\varepsilon (V^{Ai}) = i$
in the real coordinates. They satisfy the Lie algebra of a group $G$
$$
V^{Ai}\partial_i V^{Bj} - V^{Bi}\partial_i V^{Aj} = f^{ABC}V^{Cj}, \eqno (4.1)
$$
with (real) structure constants $f^{ABC}$. The metric $\gamma_{ij}$ obeys the
Killing condition
$$
\eqalignno{
{\cal L}_{V^A}\gamma_{ij} & \equiv V^{Ak}\partial_k \gamma_{ij} + \partial_i
V^{Ak}\gamma_{kj}+ (-)^{ij}\partial_j V^{Ak}\gamma_{ki} = 0,  & (4.2) \cr
{\cal L}_{V^A}\omega_{ij} & \equiv V^{Ak}\partial_k \omega_{ij} + \partial_i
V^{Ak}\omega_{kj}- (-)^{ij}\partial_j V^{Ak}\omega_{ki} = 0,  & (4.3) \cr}
$$
or equivalently
$$
\eqalignno{
{\cal L}_{V^A}\gamma^{ij} & \equiv V^{Ak}\partial_k \gamma^{ij} -
\gamma^{ik}\partial_k V^{Aj} - (-)^{(i+1)(j+1)}\gamma^{jk}\partial_k V^{Ai} =
0, & (4.4) \cr
{\cal L}_{V^A}\omega^{ij} & \equiv V^{Ak}\partial_k \omega^{ij} -
\omega^{ik}\partial_k V^{Aj} + (-)^{(i+1)(j+1)}\omega^{jk}\partial_k V^{Ai}
= 0. & (4.5) \cr} $$
{}From consistency of eqs (4.2)$\sim$(4.5) and the condition (3.1) we find
the constraints on $V^{Ai}$
$$
\partial_k V^{Al} J^k_{\ i}J^j_{\ l}   = -\partial_i V^{Aj}.
$$
In the complex coordinates this equation implies that the Killing vectors
$V^{Ai}$ are holomorphic:
$$
V^{Ai} = ({\bf R}^{Aa} ({\bf z}), \overline {\bf R}^{A \underline a}
(\overline {\bf z})).
 \eqno (4.6)
$$
Due to this property eq. (4.2) reduces to the form
$$
\partial_c ({\bf R}^{Aa}\gamma_{a \underline b})\ + \  (-)^{cb}
\partial_{\underline b} (\overline {\bf R}^{A \underline a}\gamma_
{\underline a c})
= 0   \quad \quad
 ({\rm Killing \  equation}).    \eqno (4.7)
$$
It then follows that the Killing vectors ${\bf R}^{Aa}$ and $\overline
{\bf R}^{A \underline a}$ are given by a set of real potentials  $\Sigma^A$
such that
$$
{\bf R}^{Aa} \gamma_{a \underline b}  = i \partial_{\underline b} \Sigma^A,
\quad \quad \overline {\bf R}^{A\underline a} \gamma_{\underline a b}  = -i
\partial_b \Sigma^A.   \eqno (4.8)
$$
$\Sigma^A$ are fermionic and called the Killing potentials. It is worth noting
that the isometry transformations given by the Killing vectors (4.6) can be
put in the form
$$
\eqalign{
\delta {\bf z}^a & = \epsilon^A {\bf R}^{Aa}  = \{{\bf z}^a, \epsilon^A
\Sigma^A \},  \cr
\delta \overline {\bf z}^{\underline a} & = \epsilon^A \overline {\bf R}^
{A\underline a} =
\{\overline {\bf z}^{\underline a}, \epsilon^A \Sigma^A \},  \cr} \eqno (4.9)
$$
by eqs (2.5), (3.5), (3.7), (3.8) and (4.8). Here $\epsilon^A ,  A = 1,2,
\cdots,N$
are global (real) parameters of the transformations. It is easy to show that
by these transformations the K\"ahler and Killing potentials respectively
transform  as
$$
\delta \Sigma^B  = \epsilon^A f^{ABC}\Sigma^C ,  \eqno  (4.10)
$$
and
$$
\delta K = \epsilon^A F^A ({\bf z}) + \epsilon^A \overline F^A (\overline
{\bf z}), \eqno
(4.11)
$$
with some holomorphic  functions $F^A({\bf z})$ and their complex conjugates.
Eq. (4.10) can be written  by means of the anti-bracket as
$$
\{\Sigma^A, \Sigma^B \} = f^{ABC}\Sigma^C.   \eqno (4.12)
$$
Eq. (4.10) can also be put in the form
$$
\Sigma^A = i f^{ABC}{\bf R}^{Bb}\gamma_{b\underline c}\overline {\bf R}^
{C\underline c}.
\eqno (4.13)
$$
by multiplying eq. (4.8) by $f^{ABC}$
and using  $f^{ABC}f^{ABD} = 2\delta^{CD}$. This way of calculating the Killing
potentials $\Sigma^A$
is more practical than using eq. (4.8),  if the metric
$\gamma_{a\underline b}$ is known.

\vskip 2 cm

\noindent
{\bf 5.~~The metric of the fermionic irreducible hermitian symmetric space}

The holomorphic Killing vectors ${\bf R}^{Aa}$ and $\overline {\bf R}^
{A\underline a}$ in eq. (4.6) independently satisfy the Lie-algebra (4.1),
i.e.,
$$
{\bf R}^{Ai}\partial_i {\bf R}^{Bj} - {\bf R}^{Bi}\partial_i {\bf R}^{Aj} =
f^{ABC}{\bf R}^{Cj},
$$
and the complex conjugate. These equations can be solved by
$$
{\bf R}^{Aa} = (R^{A\alpha}(z), S^{A\alpha}(z,\zeta) ), \quad \quad \quad
\alpha = 1,2,\cdots,d,
$$
with
$$
S^{A\alpha} = \zeta^\beta {\partial \over \partial z^\beta } R^{A\alpha},
$$
and $\varepsilon (R^{A\alpha})= 0=\varepsilon (S^{A\alpha}) - 1$, in
which $R^{A\alpha}$ satisfy the Lie-algebra
$$
R^{A\alpha}{\partial \over \partial z^\alpha} R^{B\beta} - R^{B\alpha}
{\partial
 \over \partial z^\alpha} R^{A\beta}
= f^{ABC}R^{C\beta}.
\eqno (5.1)
$$
These Killing vectors $R^{A \alpha}$ define a  bosonic K\"ahler manifold.
For a class of bosonic K\"ahler manifolds, called the irreducible hermitian
symmetric spaces, they
can be explicitly constructed by extending the strategy developed in ref.
15. ( The cases of   $E_6/SO(10)\otimes U(1)$ and $SU(m+n)/SU(m)\otimes
SU(n)\otimes U(1)$ have been worked out there.)
They are related to the metric of these manifolds by
$$
R^{A\alpha}\overline R^{A\underline \beta} = g^{\alpha \underline \beta},
$$
with
$$
R^{A\alpha}R^{A\beta} = 0   \quad \quad {\rm c.c.}.  \eqno (5.2)
$$
The fermionic  metric $\gamma^{ij}$ can be given in terms of these bosonic
Killing  vectors :
$$
\eqalign{
\gamma^{a \underline b} & = \left(
\matrix{ \gamma^{z\underline z} & \gamma^{z \underline \zeta} \cr
         \gamma^{\zeta \underline z} & \gamma_{\zeta \underline \zeta} \cr}
\right)
 = \left(
\matrix{ 0  &  & R^{A\alpha}\overline R^{A\underline \beta} \cr
            &  &      \cr
       R^{A\alpha}\overline R^{A\underline \beta} &   &
       R^{A\alpha}\overline S^{A\underline \beta} +
       S^{A\alpha}\overline R^{A\underline \beta} \cr}
\right)  \cr
&  \cr
\gamma^{ab} & = \gamma^{\underline a \underline b} = 0  \cr}    \eqno (5.3)
$$
in which $\gamma^{z\underline z} , \gamma^{z \underline \zeta},
\gamma^{\zeta \underline z}$ and $\gamma^{\zeta \underline \zeta}$ are
$d\times d$  matrices.
It indeed satisfies the closure property (2.7) and the Killing condition (4.4)
by means of the Lie-algebra
(4.1) and the formulae
$$
f^{ABC}R^{B\beta} R^{C\gamma} = 0, \quad \quad {\rm c.c.}.
$$
(Here recall that $\omega^{a \underline b} = i \gamma^{a \underline b}$.)
The last formulae are consequences of the condition (5.2)$^{[14]}$.
The metric (5.3) can be inverted merely by knowing the inverse of $R^{A\alpha}
\overline R^{A\underline \beta}$, denoted by $g_{\alpha \underline \beta}$:

$$
\eqalign{
\gamma_{\underline b a} & =\left(
\matrix{ \gamma_{\underline z z} & \gamma_{ \underline z \zeta} \cr
         \gamma_{\underline \zeta z} & \gamma_{\underline \zeta \zeta} \cr}
\right)  \cr
&  \cr
 & = \left(
\matrix{-g_{\underline \beta \gamma}g_{\alpha \underline \delta}
[R^{A\gamma}\overline S^{A\underline \delta} +
S^{A\gamma}\overline R^{A\underline \delta}]  &  &
g_{\underline \beta \alpha} \cr
   &   &    \cr
     g_{\underline \beta \alpha} &    &
          0   \cr}
\right)   \cr
&   \cr
 \gamma_{ba} & = \gamma_{\underline b \underline a} = 0 \cr}  \eqno (5.4)
$$
Thus we have explicitly constructed the closed 2-form (2.1). We call the
manifold with a symplectic structure given by this 2-form a fermionic
irreducible
hermitian symmetric space.
For this class of K\"ahler manifolds the Killing potentials can be
explicitly calculated by eq. (4.13).

\vskip 1 cm

 As an example we show the fermionic CP$^1$ space. It is parametrized by the
 supercoordinates $(z, \zeta)$ and their complex conjugates. The $SU(2)$
 transformations of the coordinates are given by the Killing vectors:
$$
\eqalign{
\delta z & = \epsilon^A R^{Az} = i[\epsilon^- + \epsilon^0 z -{1\over 2}
\epsilon^+z^2]  \cr
\delta \zeta &= \epsilon^A \zeta {\partial \over \partial z^\alpha} R^{Az} =
i[\epsilon^0\zeta -
\epsilon^+ z\zeta]  \cr} \eqno (5.5)
$$
and the complex conjugates\footnote \dag {We have chosen the structure
constants to be $f^{+-0}=-i$. Then the scalar product of the adjoint vectors
is given by
$a^Ab^A = a^0b^0 + a^+b^- + a^-b^+$.} .
We calculate the metric of the fermionic CP$^1$ space from eq. (5.3)
$$
\gamma^{a\underline b} = \left(
\matrix{ \gamma^{z\underline z} & \gamma^{z \underline \zeta} \cr
         \gamma^{\zeta \underline z} & \gamma^{\zeta \underline \zeta } \cr}
    \right)
= \left(
\matrix{ 0  &  (1 + {1 \over 2}z\overline z)^2    \cr
        (1 + {1 \over 2}z\overline z)^2  &
        (1 + {1 \over 2}z\overline z)(z\overline \zeta + \overline z\zeta)
        \cr}\right)  .   \eqno (5.6)
$$

\noindent
Its inverse metric is given by
$$
\gamma_{b \underline a} = \left(
\matrix{ \gamma_{\underline z  z} & \gamma_{ \underline z \zeta} \cr
         \gamma_{ \underline \zeta z} & \gamma_{\underline \zeta \zeta} \cr}
         \right)
= \left(
  \matrix{ -{z\overline \zeta + \overline z\zeta \over (1 + {1 \over 2}z
  \overline z)^3 } &  {1 \over (1 + {1 \over 2}z\overline z)^2}   \cr
   {1 \over  (1 + {1 \over 2}z\overline z)^2}  &   0     \cr}
 \right) .    \eqno (5.7)
$$

\noindent
Plugging this metric together with the Killing vectors (5.5) in eq. (4.13) we
obtain the Killing potentials
$$
\eqalign{
\Sigma^+ & = {\overline \zeta \over 1 + {1 \over 2}z\overline z } -
{\overline z (z\overline \zeta + \overline z\zeta) \over
2(1 + {1 \over 2}z\overline z)^2 } ,     \cr
\Sigma^- & = {\zeta \over 1 + {1 \over 2}z\overline z } -
{z (z\overline \zeta + \overline z\zeta) \over
2(1 + {1 \over 2}z\overline z)^2 },      \cr
\Sigma^0 & = {z\overline \zeta + \overline z \zeta \over (1 + {1 \over 2}z
\overline z)^2 }.    \cr}     \eqno (5.8)
$$

\noindent
It is worth checking that eq. (4.8) is indeed satisfied by these quantities.
The fermionic K\"ahler potential follows simply by integrating eq. (3.10) with
the metric (5.7):
$$
K = {z\overline \zeta + \overline z\zeta \over 1 + {1 \over 2}z\overline z}.
\eqno (5.9)
$$
It is a consistency check of our calculations to see that both potentials
given by (5.8) and (5.9) satisfy the properties (4.10) and (4.11) respectively.

\vskip 2 cm

\noindent
{\bf 6. The master equation}

Now we disscuss the BV formalism on the fermionic CP$^1$ space. The closed
symplectic form $\omega$ is known explicitly from the metric (5.7) by eq.
(3.7).
With this symplectic form we define the second order differential operator
according to eq. (2.9).
Then the function $\rho$ is fixed by the nilpotency condition (2.11). We find
the unique $U(1)$-invariant solution
$$
\rho = p + q{i \zeta \overline \zeta  \over (1 + {1 \over 2}z\overline z)^2 },
$$
with arbitrary constants $p(\not= 0)$ and $q$. To check this it is useful to
note that the metric (5.3) in general satisfies
$$
(-)^a\partial_a \gamma^{a \underline b} = 0 , \quad \quad \quad {\rm c.c.}.
$$
We may be interested in solving the master equation (2.12) with these
$\omega^{ij}$ and $\rho$. The solution is given by
$$
S = S_0  -({q \over 2p} + re^{-S_0}){i\zeta \overline \zeta \over (1 +
{1 \over 2}
z \overline z)^2},      \eqno (6.1)
$$
in which $S_0$ is an arbitrary function of $z$ and $\overline z$, and $r$ is
an integration constant.
We have assumed that $z$ and $\zeta$ have no space-time dependence. They can
be interpreted as coupling parameters for physical variables. Then $Z(= e^S)$
looks like the partition function of matrix models or 2-dim. topological
 conformal field theories, being a function of the coupling space. The BV
 formalism in the coupling space has been discussed by Verlinde$^{[9]}$.

  One can search for the solutions (6.1) satisfying the classical equation
$$
\{S,S\} = 0,      \eqno (6.2)
$$
which implies that $S$ is BRST invariant. Assuming reality of $S$ we find
that
$$
S = S_0 ,   \eqno (6.3)
$$
or
$$
S = a + b{i \zeta \overline \zeta  \over (1 + {1  \over 2}z\overline z)^2 },
 \eqno (6.4)
$$
with some arbitrary constants $a$ and $b$. In the language of the BRST
quantization, the first solution can be taken as a classical limit of the
full solution (6.1). Namely its BRST transformation is trivial.
By requiring the SU(2)-invariance $S_0$ is restricted to be constant.
The second solution is invariant by the $SU(2)$ transformations.

The master equation (2.12) may be solved also by allowing $z$ and $\zeta$
space-time dependence. In this case the first piece of the equation suffers
from the singularity
$\delta (0)$. An appropriate  regularization is necessary. It is not the aim
of this letter to discuss
regularization of this singularity. Therefore we study the classical master
equation (6.2),
which is still of great interest. Remarkably there is a solution for the
general K\"ahler group manifold discussed above, although it would not be the
unique one.  It is given by
the action of a 2-dim. field theory
$$
S = \int d^2 x G\partial_- \Sigma^0.  \eqno (6.5)
$$
Here $G$ is an arbitrary $U(1)$-invariant function of bosons $z^\alpha
(x^+,x^-)$ , chiral fermions $\zeta^\alpha (x^+,x^-)$ and their complex
coordinates. $\Sigma^0$
is the $U(1)$-component of the Killing potentials given by eq. (4.13). We
can verify that this action satisfies eq. (6.2) by calculating
$$
\eqalign{
\{S,S\} & = i(-)^a\partial_a S \gamma^{a\underline b}\partial_{\underline b}S
 - i(-)^a\partial_{\underline a} S \gamma^{\underline a b}\partial_b S    \cr
  & = 2\int d^2x [-i \partial_a G \gamma^{a \underline b}\partial_{\underline
 b}\Sigma^0 + i \partial_{\underline a} G \gamma^{\underline a b}\partial_b
 \Sigma^0 ]\partial_- G \partial_- \Sigma^0   \cr
 & = -2\int d^2x [{\bf R}^{0a}\partial_a G + \overline {\bf R}^{0 \underline a}
 \partial_{\underline a} G] \partial_- G\partial_- \Sigma^0    \cr
 & = 0\ ,    \cr}
$$
using eq. (4.8) and $U(1)$-invariance of $G$.
In the CP$^1$ case the solution (6.5) can be written in the general form
$$
S = \int d^2 x [(z \overline \zeta + \overline z \zeta)f
               + i(z \overline \zeta - \overline z \zeta)g]
               \partial_- [{z \overline \zeta + \overline z \zeta \over
               (1 + {1\over 2}z \overline z)^2} ],
$$
in which $f$ and $g$ are arbitrary real functions of $z\overline z$. The BRST
transformations of the fields are given by $\{z, S\}$ and $\{\zeta, S\}$ ,
so that
$$
\eqalign{
\delta_{BRST} (z \overline \zeta + \overline z \zeta)
& = -2z\overline z (1 + {1\over 2}z \overline z) (z \overline \zeta +
\overline z \zeta)g \partial_- \Sigma^0 ,    \cr
\delta_{BRST} (z \overline \zeta - \overline z \zeta)
& = -2i\ z\overline z (1 + {1\over 2}z \overline z) [(z \overline \zeta +
\overline z \zeta)f \cr
& \quad  + (1 + {1 \over 2}z \overline z)\{(z\overline \zeta + \overline z
\zeta)f' + i(z\overline \zeta - \overline z \zeta)g'\}] \partial_- \Sigma^0 ,
\cr
\delta_{BRST} (z \overline z)
& =  -2z\overline z (1 + {1\over 2}z \overline z)^2 g \partial_-\Sigma^0. \cr}
$$
Indeed the action (6.5) is invariant by these BRST transformations.

\vskip 1 cm

\noindent
{\bf 7.~~Conclusions}

An explicit expression of the closed fermionic 2-form $\omega$ is of primary
interest to start  the  BV formalism. Hence the supermanifold with such a
symplectic structure  should be properly understood.
In this note we have discussed how  the K\"ahler structure can be
incorporated in the supermanifold.  The closed fermionic form $\omega$  is
then reduced to the K\"ahler 2-form  . It has been given by means of the
fermionic K\"ahler potential.
Thus  a fermionic K\"ahler manifold appears as the natural phase space of
the BV formalism.  Secondly we have introduced an iso\-metry to this
manifold.
The novelty obtained by doing this was that the fermionic K\"ahler 2-form
$\omega$ is given by the Killing vectors of the isometry. The existence of
the Killing potentials for the isometry is most characteristic in the K\"ahler
group manifold.
They are fermionic in our case. We have given the formula to construct
them out of
the Killing vectors, eq. (4.13). It holds for the general K\"ahler group
manifold.  Finally we have studied the master equation with the symplectic
structure given above. The space-time independent solution has been given by
eq. (6.1).
The BRST invariant solution (6.4) happens to be  $SU(2)$-invariant
as well.
For space-time dependent solutions we have studied the classical
master equation.
It is solved by eq. (6.5) for the irreducible fermionic hermitian symmetric
space. The solution has merely $U(1)$ symmetry. This could be just one of
many other solutions. The authors have not yet succeeded in finding a solution
which is invariant by the
full isometry.
Space-time dependent solutions are  certainly more interesting from the
physical point of view. It deserves further study in order to find new
solutions.

\vskip 2cm

\noindent
{\bf Acknowledgments}
We are grateful to A. Van Proeyen for illuminating discussions. One of the
authors (S.A) thanks the Research Council of K.U. Leuven for the financial
support.

\vskip 1 cm

\noindent
{\bf References}

\noindent
\item{1.} J. Zinn-Zustin, Springer Lecture Notes in Physics 37(1975)1.
\item{2.} I.A. Batalin and G. A. Vilkovisky, Phys. Lett. B102(1981)27; Phys.
Rev. D28(1983)2567.
\item{3.} R. Kallosh, Phys. Lett. B195(1987)369.
\item{4.} B. de Wit and J. W. van Holten, Phys. Lett. B79(1979)389;
\item{5.} E. Witten, Mod. Phys. Lett. A5(1990)487.
\item{6.} A. Schwarz, ``Geometry of Batalin-Vilkovisky quantization", UC Davis
pre\-print, hep-th/9205088, May 1992;
\item{} I. A. Batalin and I. V. Tyutin, ``On possible generalizations of
field-antifield formalism", FIAN/TD/18-92, November 1992.
\item{7.} E. Witten, ``On background independent open-string field theory",
IASSNS-HEP-92/53, hep-th/9208027, August 1992.
\item{8.} H. Hata and B. Zwiebach, ``Developing the covariant
Batalin-Vilkovisky
approach to string theory", MIT-CTP-2177, hep-th/9301097, January 1993.
\item{9.} E. Verlinde, Nucl. Phys. B381(1992)141.
\item{10.} J. Bagger and E. Witten, Phys. Lett. B118(1982)103.
\item{11.} Helgason, ``Differential geometry, Lie groups and symmetric
spaces",
Academic Press, N-Y, 1978.
\item{12.} W. Troost, P. van Nieuwenhuizen and A. Van Proeyen, Nucl. Phys.
B\-333  (1990)\-727;
\item{} A. Van Proeyen and S. Vandoren, in preparation.
\item{13.} O.M. Khudaverdian and A.P. Nersessian, ``On the geometry of the BV
formalism", UGVA-93/03-807, March 1993.
\item{14.} S. Aoyama, Z. Phys. C32(1986)113.
\item{15.} Y. Achiman, S. Aoyama and J. W. van Holten, Nucl. Phys. B258(1985)
179; Phys. Lett. B141(1984)64.

\vskip 2cm

\bye